\documentclass[10pt]{iopart}

\usepackage{iopams}
\usepackage{graphicx}
\usepackage{cite}

\newcommand{\prd}{Phys. Rev. D}

\usepackage{bm}

\newcommand{\be}{\begin{equation}}
\newcommand{\ee}{\end{equation}}
\newcommand{\ba}{\begin{eqnarray}}
\newcommand{\ea}{\end{eqnarray}}
\newcommand{\<}{\langle}
\renewcommand{\>}{\rangle}

\newcommand{\del}{\nabla}
\newcommand{\li}{&}
\newcommand{\sfrac}{\frac}
\newcommand{\aperp}{a_{\perp }}

\newcommand{\apar}{a_{\parallel}}
\newcommand{\kapr}{\kappa(r)r^2}
\newcommand{\Hperp}{H_{\perp}}
\newcommand{\Hperpo}{H_{\perp_0}}
\newcommand{\Hpar}{H_{\parallel}}
\newcommand{\omegmo}{\Omega_{{m}}}
\newcommand{\omegko}{\Omega_{{\kappa}}}

\renewcommand{\;}[2]{{\frac{#1}{#2}}}
\newcommand{\udot}{A}
\newcommand{\sdel}{\tilde\nabla}

\newcommand{\gmgeta}{\eta}
\newcommand{\gmgpsi}{\varsigma}
\newcommand{\gmgchi}{\chi}

\newcommand{\gmgk}{\varphi}
\newcommand{\gmgalpha}{v}

\newcommand{\gmggamma}{w}
\newcommand{\gmgomega}{\Delta}
\renewcommand{\H}{\mathcal{H}}

\newcommand{\T}{{\text{\tiny\it (T)}}}
\newcommand{\TF}{{\text{\tiny\it (TF)}}}
\newcommand{\GI}{{\text{\tiny\it (GI)}}}

\newcommand{\p}{\partial}

\newcommand{\dec}{{*}}
\newcommand{\eq}{{\text{eq}}}

\newcommand{\text}{\mathrm}

\newcommand{\D}{\tilde{\nabla}}

\newcommand{\ep}{\varepsilon}
\renewcommand{\a}{\mu}
\renewcommand{\e}{\rho}
\newcommand{\f}{\eta}

\begin{document}

\title{Inhomogeneity and the foundations of concordance cosmology}

\author{Chris Clarkson$^1$ and Roy Maartens$^{1,2}$}

\address{$^1$Centre for Astrophysics, Cosmology \& Gravitation, and, Department of Mathematics and
Applied Mathematics, University of Cape Town, Rondebosch 7701, Cape Town, South Africa\\ $^2$ Institute of Cosmology \& Gravitation, University of Portsmouth, Portsmouth~PO1~3FX, UK\footnote{Permanent address}}

\ead{\mailto{chris.clarkson@uct.ac.za}, \mailto{roy.maartens@port.ac.uk}}

\begin{abstract}

The apparent accelerating expansion of the Universe is forcing us to examine the foundational aspects of the standard model of cosmology -- in particular, the fact that dark energy is a direct consequence of the homogeneity assumption.
We discuss the foundations of the assumption of spatial homogeneity, in the case when the Copernican Principle is adopted. 
We present results that show how (almost-) homogeneity follows from (almost-) isotropy of various observables.
The analysis requires the fully nonlinear field equations -- i.e., it is not possible to use second- or higher-order perturbation theory, since one cannot assume a homogeneous and isotropic background. Then we consider what happens if the Copernican Principle is abandoned in our Hubble volume. The simplest models are inhomogeneous but spherically symmetric universes which do not require dark energy to fit the distance modulus.  Key problems in these models are to compute the CMB anisotropies and the features of large-scale structure. We review how to construct perturbation theory on a non-homogeneous cosmological background, and discuss the complexities that arise in using this to determine the growth of large-scale structure.

\end{abstract}

\vspace*{-.5cm}


\section{INTRODUCTION}

The standard model of the Universe -- the LCDM ``concordance" model -- is a perturbed FLRW model containing cold dark matter and dark energy in the form of a cosmological constant $\Lambda$. This model is highly successful in being able, up to now, to fit all cosmological observations, with the same small set of parameters~\cite{1}. However, there is as yet no satisfactory explanation for the value of $\Lambda$, which appears in Einstein's field equations in precisely the same form as the total vacuum energy of quantum fields. This term is responsible for the late-time acceleration of the expansion of the Universe within the spatially homogeneous FLRW framework.

The unresolved nature of $\Lambda$ and of alternative forms of dark energy throws into sharp focus the foundations of the standard model -- in particular, the spatial homogeneity assumption. We will address two questions in this paper:

\begin{itemize}
\item[{\bf (Q1)} ] What is the basis for the spatial homogeneity assumption?
\item[{\bf (Q2)} ] If we drop this assumption, can we find a model that is consistent with all cosmological observations?
\end{itemize}

In summary, the current answer to (Q1) is that there are various results which motivate homogeneity (assuming the Copernican Principle); the most important one is that
the observed (almost-)isotropy of the CMB provides strong evidence for (almost-)homogeneity of the Universe on large scales~\cite{2}. But crucial open questions remain, which we identify and discuss.
Standard cosmological perturbation theory is powerless here -- the argument depends on a fully nonlinear analysis, since one cannot assume an FLRW background.

The current answer to (Q2) is that the simplest inhomogeneous cosmological models -- isotropic LTB (Lema\^\i tre-Tolman-Bondi) models without any dark energy, and with our galaxy at the centre of a large void~-- are able to fit current supernova and some other data~\cite{MT1,MT2,MHE,PascualSanchez:1999zr,celerier1,Tomita:2000jj,IKN,Mof1,Mof2,celerier2,VFW,CR,EM,BMN,bolejko,enqvist,YKN,AAG,Alnes:2006pf,alnes,ABNV,celerier3,Romano:2007zz,gbh1,gbh2,Enqvist,ZMS,CFL,BW,CCF,CFZ,CBKH,FLSC,mortsell,Mof3,Yoo:2010qy,Romano:2009ej,Romano:2009mr,Romano:2009qx}. In this framework, the late-time acceleration becomes a mis-interpretation of an effect due to nonlinear inhomogeneity and curvature. In order to test these models against the full range of cosmological observations, it is necessary to develop  perturbation theory on an LTB background. Although standard cosmological perturbation theory provides some qualitative guidance, perturbations on an inhomogeneous background are fundamentally different -- in particular, there is no longer a simple separation of scalar, vector and tensor modes at first order. We review recent developments on this~\cite{CCF}, and discuss remaining open questions.

\section{WHAT IS THE BASIS FOR SPATIAL HOMOGENEITY? }

Isotropy is directly observable in principle, and indeed we have excellent data to show that the CMB is isotropic about us to within one part in $\sim 10^{5}$ (once the dipole is interpreted as due to our motion relative to the cosmic frame, and removed by a boost). Observations of the galaxy distribution do not have the same precision, but there is no evidence for anisotropy. However, spatial homogeneity {\em cannot} be directly observed -- for the simple reason that we are effectively unable to move away from here-and-now in order to probe spatial variations on constant time slices. The nature of cosmic time and length scales means that effectively our observations only access the past lightcone of here-and-now. Direct observation therefore cannot distinguish between a homogeneous distribution of matter that evolves with time down the past lightcone, and inhomogeneity with a different time evolution -- since the past light cone only accesses a 2-sphere in each constant-time slice.

The best that we can do currently is to assume a Copernican Principle -- i.e., that our galaxy does not have an atypical view of the Universe. If we adopt the Copernican assumption -- so that all galactic observers are assumed to see a near-isotropic CMB -- then a nonlinear analysis gives evidence for near-homogeneity on cosmological scales, as we describe below. The fully nonlinear field and Boltzmann equations in covariant Lagrangian form are summarized for convenience in Appendix~A.

We first discuss the exact results for the case of an exactly isotropic CMB. Then we consider the almost-isotropic case. Finally we look at what can be deduced if observations of galaxies down the past light cone are isotropic.

\subsection{{\bf Exactly isotropic CMB}}

What can we say if the CMB is exactly isotropic for fundamental observers? The pioneering result is due to Ehlers, Geren and Sachs (1968)~\cite{EGS}. The original EGS work assumed that the only source of the gravitational field was the radiation, i.e., they neglected matter (and they had $\Lambda=0$). We generalise their result to include self-gravitating matter and dark energy (extending~\cite{2,TreEll71,FerMorPor99,CB,CC,Rasanen:2009mg}):

\begin{quote}
\textbf{\em (EGS) Radiation isotropy $\rightarrow$ FLRW}\\
In a region, if
\begin{itemize}
\item collisionless radiation is exactly isotropic,
\item the radiation four-velocity is geodesic and expanding,
\item there is dust matter and dark energy in the form of $\Lambda$, quintessence or a perfect fluid,
\end{itemize}
then the metric is FLRW in that region.
\end{quote}

The fundamental 4-velocity $u^\a$ is the radiation 4-velocity, which has zero 4-acceleration and positive expansion:
\begin{equation}\label{av}
A_\a=0\,,~~\Theta >0\,.
\end{equation}
Isotropy of the radiation distribution about $u^\a$ means that (individual) photon peculiar velocities are isotropic for comoving observers; thus in momentum space, the photon distribution depends only on components of the 4-momentum $p^\a$ along $u^\a$, i.e., on the photon energy $E=-u_\a p^\a$:
\begin{equation}\label{iso}
f(x,p)=F(x,E),~~F_{\a_1 \cdots \a_\ell}=0~\mbox{for}~ \ell \geq 1\,.
\end{equation}
In other words, all covariant multipoles of the distribution function beyond the monopole, defined in Eq.~(\ref{r3}), must vanish. In particular, as follows from Eq.~(\ref{em3}), the momentum density (from the dipole) and anisotropic stress (from the quadrupole)  vanish:
\begin{equation}
\label{qpv}
q_r^\a=0=\pi_r^{\a\nu}\,.
\end{equation}
Equation~(\ref{iso}) also implies that the radiation brightness octopole $\Pi_{\a\nu\alpha}$ and hexadecapole $\Pi_{\a\nu\alpha\beta}$ are zero. These are source terms in the anisotropic stress evolution equation, which is the $\ell=2$ case of Eq.~(\ref{r26}). In general fully nonlinear form, the $\pi_r^{\a\nu}$ evolution equation is
\begin{eqnarray}
&&\dot{\pi}_{r}^{\langle \a\nu \rangle}+{{4\over3}}\Theta
\pi_{r}^{\a\nu } +{{8\over15}}\rho_{r}\sigma^{\a\nu }+
{{2\over5}}\D^{\langle \a}q_{r}^{\nu\rangle} +2 A^{\langle \a} q_{r}^{\nu\rangle} -2\omega^{\alpha}\ep_{\alpha \beta }{}{}^{(\a}
\pi_{r}^{\nu) \beta} ~~~~~
\nonumber\\
&& ~~~~~~~~  +{{2\over7}}\sigma_{\alpha}{}^{\langle
\a}\pi_{r}^{\nu\rangle \alpha} +{8\pi\over35}\D_{\alpha}
\Pi^{\a\nu \alpha}  -{32\pi\over315} \sigma_{\alpha \beta } \Pi^{\a\nu \alpha \beta }
= 0. \label{nl8}
\end{eqnarray}
Isotropy removes all terms on the left except the third, and thus enforces a shear-free expansion of the fundamental congruence:
\begin{equation}\label{sv}
\sigma_{\a\nu}=0\,.
\end{equation}

We can also show that $u^\a$ is irrotational as follows. Together with Eq.~(\ref{av}), momentum conservation for radiation, i.e., Eq.~(\ref{e3i}) with $I=r$, reduces to
\begin{equation}\label{mv}
\D_\a \rho_r=0\,.
\end{equation}
Thus the radiation density is homogeneous relative to fundamental observers.
Now we invoke the exact nonlinear identity for the covariant curl of the gradient, Eq.~(\ref{ri1}):
\begin{equation}\label{}
\mbox{curl}\, \D_\a \rho_r = - 2\dot \rho_r \omega_\a~ \Rightarrow~ \Theta \rho_r \omega_\a =0\,,
\end{equation}
where we have used the energy conservation equation~(\ref{e1i}) for radiation. By assumption $\Theta >0$, and hence we deduce that the vorticity must vanish:
\begin{equation}\label{vv}
\omega_\mu =0\,.
\end{equation}
Then we see from the curl shear constraint equation~(\ref{c3}) that the magnetic Weyl tensor must vanish:
\begin{equation}\label{hv}
H_{\a\nu}=0\,.
\end{equation}

Furthermore, Eq.~(\ref{mv}) actually tells us that the expansion must also be homogeneous. From the radiation energy conservation equation~(\ref{e1i}), and using Eq.~(\ref{qpv}), we have $\Theta=-3{\dot\rho_r}/{4\rho_r}$.
On taking a covariant spatial gradient and using the commutation relation Eq.~(\ref{timespace}), we find
\be \label{tv}
\D_\a\Theta=0\,.
\ee
Then the shear divergence constraint, Eq.~(\ref{c2}), enforces the vanishing of the total momentum density in the fundamental frame,
\begin{equation}\label{qv}
q^\a \equiv \sum_I q_I^\a =0~ \Rightarrow ~\sum_I \gamma_I^2(\rho^*_I + p^*_I)v_I^\a=0  \,.
\end{equation}
The second equality follows from Eq.~(\ref{t6}), using the fact that the baryons, CDM and dark energy (in the form of quintessence or a perfect fluid) have vanishing momentum density and anisotropic stress in their own frames, i.e.,
\begin{equation}\label{perf}
q_I^{*\a}=0= \pi_I^{*\a\nu}\,,
\end{equation}
where the asterisk denotes the intrinsic quantity (see Appendix A). If we include other species, such as neutrinos, then the same assumption applies to them. Except in artificial situations, it follows from Eq.~(\ref{qv}) that
\begin{equation}\label{pvan}
v_I^\a=0\,,
\end{equation}
i.e., the bulk peculiar velocities of matter and dark energy [and any other self-gravitating species satisfying Eq.~(\ref{perf})] are forced to vanish -- all species must be comoving with the radiation.

The comoving condition~(\ref{pvan}) then imposes the vanishing of the total anisotropic stress in the fundamental frame:
\begin{equation}\label{piv}
\pi^{\a\nu}\equiv \sum_I \pi_I^{\a\nu} =\sum_I \gamma_I^2(\rho^*_I + p^*_I)v_I^{\langle \a}v_I^{\nu \rangle}=0  \,,
\end{equation}
where we used Eqs.~(\ref{t7}), (\ref{perf}) and (\ref{pvan}). Then the shear evolution equation~(\ref{e5}) leads to a vanishing electric Weyl tensor
\begin{equation}\label{evan}
E_{\a\nu}=0\,.
\end{equation}
Equations~(\ref{qv}) and (\ref{piv}), now lead via the total momentum conservation equation~(\ref{e3}) and the $E$-divergence constraint~(\ref{c4}), to homogeneous total density and pressure:
\begin{equation}\label{rpv}
\D_\a\rho = 0 = \D_\a p\,.
\end{equation}

Equations~(\ref{av}), (\ref{sv}), (\ref{hv}), (\ref{tv}), (\ref{qv}), (\ref{piv}) and (\ref{rpv}) constitute a covariant characterization of an FLRW spacetime. This establishes the EGS result, generalized from the original to include self-gravitating matter and dark energy. (We have also provided an alternative, 1+3 covariant, analysis.) It is straightforward to include other species such as neutrinos. The critical assumption needed for all species is the vanishing of the intrinsic momentum density and anisotropic stress, i.e., Eq.~(\ref{perf}). Equivalently, the energy-momentum tensor for the $I$-component should have perfect fluid form in the $I$-frame (we rule out a special case that allows unphysical total anisotropic stress~\cite{CC}). The isotropy of the radiation and the geodesic nature of its 4-velocity then enforce the vanishing of (bulk) peculiar velocities $v_I^\a$. We emphasize that one does {\em not} need to assume that the matter or other species are comoving with the radiation -- it follows from the assumptions on the radiation.\\

The original EGS result (i.e., without matter or dark energy) was generalized by Ellis, Treciokas and Matravers (1983)~\cite{Ellis:1983} to a much weaker assumption on the photon distribution: only the dipole, quadrupole and octopole need be assumed zero. The key step is to show that the shear vanishes, without having zero hexadecapole -- the quadrupole evolution equation~(\ref{nl8}) no longer automatically gives $\sigma_{\a\nu}=0$, and we need to find another way to show this. The ETM trick is to return to the Liouville multipole equation~(\ref{r25}). The $\ell=2$ multipole of this equation, with $F_\a=F_{\a\nu}= F_{\a\nu\alpha}=0$, gives
\begin{equation}\label{}
{12 \over 63}{\p \over \p E}\left( E^5\sigma^{\a\nu}F_{\a\nu\alpha\beta} \right)+ E^5 {\p F\over \p E}\,\sigma_{\alpha\beta}=0\,.
\end{equation}
We integrate over $E$ from 0 to $\infty$, and use the convergence property $E^5 F_{\a\nu\alpha\beta} \to 0$ as $E \to \infty$. This gives
\begin{equation}\label{}
\sigma_{\alpha\beta} \int_0^\infty E^5 {\p F\over \p E}\mathrm{d} E=0\,.
\end{equation}
Integrating by parts, the integral reduces to $-5\int_0^\infty E^4 F \mathrm{d} E$. Since $F>0$, the integral is strictly negative, and thus we arrive at vanishing shear, $\sigma_{\a\nu}=0$. Then our proof above proceeds as before.

Thus we can present a stronger generalization of the EGS result:

\begin{quote}
\textbf{\em (EGS-ETM) Radiation partial isotropy $\rightarrow$ FLRW}

In a region, if
\begin{itemize}
\item collisionless radiation has vanishing dipole, quadrupole and octopole,
\begin{equation}\label{}
F_\a= F_{\a\nu}= F_{\a\nu\alpha}=0\,,
\end{equation}
\item the radiation four-velocity is geodesic and expanding,
\item there is dust matter and dark energy in the form of $\Lambda$, quintessence or a perfect fluid,
\end{itemize}
then the metric is FLRW in that region.
\end{quote}

In summary, {\em the EGS-ETM result, suitably generalized to include baryons and CDM and dark energy, is the most powerful basis that we have -- within the framework of the Copernican Principle -- for background spatial homogeneity and thus an FLRW background model.}

Although this result applies only to the ``background Universe", its proof nevertheless requires a fully nonperturbative analysis.

\subsection{{\bf The real universe: almost-isotropic CMB}}

In practice we can only observe approximate isotropy. Is the EGS result stable -- i.e., does almost-isotropy of the CMB lead to an almost-FLRW Universe? This would be the {\em realistic} basis for a spatially homogeneous Universe (assuming the Copernican Principle). It  was shown to be the case, subject to further assumptions, by Stoeger, Maartens and Ellis~\cite{2,Maartens:1994qq}:

\begin{quote}
\textbf{\em (Almost-EGS) Realistic basis for spatial homogeneity}

In a region of an expanding Universe with cosmological constant, if all observers comoving with the matter measure an almost isotropic distribution of collisionless radiation, and if some of the time and spatial derivatives of the covariant multipoles are also small, then the region is almost homogeneous.
\end{quote}

We emphasize that the perturbative assumptions are purely on the photon distribution, not the matter or the metric -- and one has to prove that the matter and metric are then perturbatively close to FLRW. Once again, a nonperturbative analysis is essential, since we are trying to prove an almost-FLRW spacetime, and we cannot assume it a priori.

Almost-isotropy of the photon distribution means that
\begin{equation}\label{}
F_{\a_1 \cdots \a_\ell}(x,E)= {\cal O}(\epsilon), ~~ \ell\geq 1\,,
\end{equation}
where $\epsilon$ is a (dimensionless) smallness parameter. 
The brightness multipoles $\Pi_{M_\ell}$ have dimensions of energy density and we therefore normalize them to the monopole $\Pi= \rho_r/4\pi$:
\begin{equation}\label{}
{\Pi_{M_\ell} \over \Pi}= {\cal O}(\epsilon)\,.
\end{equation}

The task is to show that the relevant kinematical, dynamical and curvature quantities, suitably non-dimensionalized, are ${\cal O}(\epsilon)$. For example, the dimensionful kinematical quantities may be normalized by the expansion, $\sigma_{\a\nu}/\Theta, \omega_\a/\Theta$. The proof then follows the same pattern as our proof above of the exact EGS result -- except that at each stage, we need to show that quantities are ${\cal O}(\epsilon)$ rather than equal zero.

However, in order to show this, we need smallness not just of the multipoles, but also of some of their derivatives. Smallness of the multipoles does not directly imply smallness of their derivatives, and we have to assume this~\cite{Rasanen:2009mg,Rasanen:2008be}. It has been claimed~\cite{Rasanen:2009mg,Rasanen:2008be} that the derivatives are generically {\em not} small, but this has not been rigorously established -- which is not surprising, since it is a difficult problem and may be related to the averaging problem in cosmology. The issue needs to be settled by further analysis.

If all observers measure small multipoles, then it may be possible to show -- perhaps using other observations --  that the time and space derivatives on cosmologically significant scales must also be small. A number of experiments has been proposed to test the Copernican Principle by looking for violations of isotropy at events down our past lightcone. These include looking for spectral distortions of CMB photons scattered by ionized gas~\cite{goodman,CS}:  such distortions are induced by anisotropies in the CMB as seen by distant observers, and so provide in principle a neat way of confirming the Copernican assumption as used here. A similar test uses the kinematic Sunyaev-Zeldovich effect in clusters to observe the dipole around distant observers~\cite{gbh2}. The almost-EGS result then gives a framework for probing inhomogeneities via such observations. Indeed, these tests may provide a way of constraining spatial gradients of the low-$\ell$ multipoles~\cite{ccrm}.

In addition, it may be possible to strengthen the almost-EGS result above by proving that it is sufficient for only the first 3 multipoles and their derivatives to be small. This would be an almost-EGS-ETM result, and would represent a more realistic foundation for almost-homogeneity than the almost-EGS result.

Note also that the almost-EGS result has not been proven for quintessence or perfect fluid dark energy, and this needs further investigation~\cite{ccrm}.

\subsection{{\bf Isotropic matter distribution on the past lightcone}}

We have seen the power of an (almost-) isotropic CMB as the fundamental basis for a spatially homogeneous background model of the Universe. What can we say if the matter distribution is (almost-) isotropic on the past lightcone of the observer?

A ``matter" counterpart to the EGS result is based on analysis of the observations of galaxies down the past lightcone of the observer. Without assuming the Copernican Principle, we have the following isotropy result~\cite{7,7a,7b}
\begin{quote}
\textbf{\em Matter lightcone-isotropy $\rightarrow$ spatial isotropy}

If one fundamental observer comoving with the matter measures isotropic area distances, number counts, bulk peculiar velocities, and lensing, in an expanding dust Universe with $\Lambda$, then the spacetime is isotropic about the observer's worldline.
\end{quote}

The original result neglected $\Lambda$ and CDM, and we can incorporate both in the same way as in the EGS case. Note that isotropy of (bulk) peculiar velocities seen by the observer is equivalent to vanishing proper motions (tranverse velocities) on the observer's sky. Isotropy of lensing means that there is no distortion of images, only magnification.

The proof of this result requires a non-perturbative approach -- there is no background to perturb around. Since the data is given on the past lightcone of the observer, the covariant Lagrangian approach used in the previous subsection is not suitable. Instead, we need the fully general metric, adapted to the past lightcones of the observer worldline ${\cal C}$. We define observational coordinates $x^\a=(w,y,\theta,\phi)$, where $x^a=(\theta,\phi)$ are the celestial coordinates, $w=\,$const are the past light cones on ${\cal C}$ ($y=0$), normalized so that $w$ measures proper time along ${\cal C}$, and $y$ measures distance down the light rays $(w,\theta,\phi)=\,$const. A convenient choice for $y$ is $y=z$ (redshift) on the lightcone of here-and-now, $w=w_0$, and then keep $y$ comoving with matter off the initial lightcone, so that $u^y=0$. Then the matter 4-velocity and the photon wave-vector are
\begin{equation}\label{}
u^\a=(1+z) (1,0, V^a)\,,~~ k_\a=w_{,\a}\,, ~~ 
1+z=u_\a k^\a,
\end{equation}
where $V^a=\mathrm{d} x^a/\mathrm{d} w$ are the transverse velocity components on the observer's sky.

The metric is
\begin{eqnarray}
\mathrm{d} s^2 &=&\! -A^2\mathrm{d} w^2+ 2B\mathrm{d} w \mathrm{d} y+ 2C_a\mathrm{d} x^a \mathrm{d} w+D^2(\!\mathrm{d}\Omega^2+ L_{ab}\mathrm{d} x^a \mathrm{d} x^b )~~ \label{} \\
A^2 &=& (1+z)^{-2}+2C_aV^a+ g_{ab}V^a V^b\,,~~~ B={\mathrm{d} v \over \mathrm{d} y},
\end{eqnarray}
where $D$ is the area distance, and $L_{ab}$ determines the lensing distortion of images via the shear of lightrays,
\begin{equation}\label{}
\hat\sigma_{ab}= {D^2 \over 2B} {\p L_{ab} \over \p y}.
\end{equation}
The number of galaxies in a solid angle $\mathrm{d}\Omega$ and a null distance increment $\mathrm{d} y$ is
\begin{equation}\label{}
\mathrm{d} N = Sn(1+z)D^2B \mathrm{d}\Omega \mathrm{d} y\,,
\end{equation}
where $S$ is the selection function and $n$ is the number density.

Before specializing to isotropic observations, we identify how the observations in general and in principle determine the geometry of the past light cone $w=w_0$ of here-and-now, where $y=z$:
\begin{itemize}
\item
Area distances directly determine the metric function $D(w_0,z,x^a)$.
\item
The number counts (given a knowledge of $S$) determine $Bn$ and thus, assuming a knowledge of the bias, they determine $\mu_{m}(w_0,z,x^a)\equiv B(w_0,z,x^a) \rho_m(w_0,z,x^a)$, where $\rho_m=\rho_b+\rho_c$ is the total matter density.
 \item
Transverse (proper) motions in principle directly determine $V^a(w_0,z,x^b)$.
\item
Image distortion determines $L_{ab}(w_0,z,x^c)$. (The differential lensing matrix $\hat\sigma_{ab}$ is determined by $L_{ab}, D,B$.)
\end{itemize}
Then~\cite{7,7a}:
\begin{quote}
\textbf{\em Lightcone observations $\rightarrow$ spacetime metric}

Observations $(D, N, V^a,L_{ab})$ on the past lightcone $w=w_0$ determine in principle $(g_{ab},u^\a,\mu_m\equiv B\rho_m)$ on the lightcone. This is exactly the information needed for Einstein's equations to determine $B,C_a$ on $w=w_0$, so that the metric and matter are fully determined on the lightcone. Finally, the past Cauchy development of this data determines $g_{\a\nu},u^\a,\rho_m$ in the interior of the past lightcone.
\end{quote}

If we assume that observations are isotropic, then
\begin{equation}\label{misob}
{\p D \over \p x^a}= {\p N\over \p x^a}=V^a=L_{ab}=0\,.
\end{equation}
Momentum conservation  and the $yy$ field equation then give the following equations on $w=w_0$~\cite{7,7b}:
\begin{eqnarray}
C_a &=& (1+z)^{-1}\int_0^z(1+z)B_{,a} dz \\
B &=&{\mathrm{d} v \over \mathrm{d} z} =  2D'\left[2-\int_0^z(1+z)^2D\mu_m dz\right]^{-1},
\end{eqnarray}
where a prime denotes $\p/\p z$. These imply that $B_{,a}=0=C_a$, so that $\rho_{m,a}=0$ -- and hence the metric and matter are isotropic on $w=w_0$. This can only happen if the interior of $w=w_0$ is isotropic. If observations remain isotropic along ${\cal C}$, then the spacetime is isotropic.

Adopting the Copernican Principle, it follows that all observers see isotropy and therefore that spacetime is isotropic about all galactic worldlines -- and hence spacetime is FLRW. This result then becomes a ``matter" alternative to the EGS, as a basis for FLRW:

\begin{quote}
\textbf{\em Matter lightcone-isotopy $\rightarrow$ FLRW}

In an expanding dust region with $\Lambda$, if all fundamental observers measure isotropic area distances, number counts, bulk peculiar velocities, and lensing, then the spacetime is FLRW.
\end{quote}

In essence, this is the Cosmological Principle, but derived from observed isotropy and not from assumed spatial isotropy. Note the significant number of observable quantities required.

We can actually give a much stronger statement than this, based only on distance data.
An important though under-recognized theorem due to Hasse and Perlick tells us that~\cite{HP}

\begin{quote}
\textbf{\em (HP) Isotropic distances $\rightarrow$ FLRW}

If all fundamental observers in an expanding spacetime region measure isotropic area distances up to third-order in a redshift series expansion, then the spacetime is FLRW in that region.
\end{quote}

The proof of this relies on performing a series expansion of the distance-redshift relation in a general spacetime, using the method of Kristian and Sachs~\cite{KS}, and looking at the spherical harmonic multipoles order by order. We illustrate the proof in the 1+3 covariant approach (in the case of zero vorticity). Performing a series expansion in redshift of the distance modulus, we have, in the notation of~\cite{MacEllis},
\ba\label{m-z-general}
m- M-25&=&5\log_{10}z-5\log_{10}
\left.K^\a K^\nu\del_\a u_\nu\right|_o+
\nonumber\\&&
\sfrac52\log_{10}e\Bigg\{
\Bigg[4-\frac{K^\a K^\nu K^\lambda\del_\a\del_\nu u_\lambda}{(K^\kappa K^\e\del_\kappa u_\e)^2}
\Bigg]_Oz
\nonumber\\
 &&-\Bigg[2+\frac{R_{\a\nu}K^\a K^\nu}{6(K^\lambda K^\kappa\del_\lambda u_\kappa)^2}-
\frac{3(K^\a K^\nu K^\lambda\del_\a\del_\nu u_\lambda)^2}{4(K^\kappa K^\e\del_\kappa u_\e)^4}
\nonumber\\&&
+
\frac{K^\a K^\nu K^\lambda K^\kappa\del_\a\del_\nu\del_\lambda u_\kappa}{3(K^\e K^\f\del_\e u_\f)^3}\Bigg]
_Oz^2
\Bigg\}+{\cal
O}(z^3),
\ea
where
\be
K^\a=\frac{k^\a}{\left.u^\nu k_\nu\right|_O}~~~~\hbox{and}~~~~
\left.K^\a\right|_O=\left.-(u^\a+e^\a)\right|_O,
\ee
denotes a past-pointing null vector at the observer $O$ in the direction $e^\a$.
Comparing with the standard FLRW series expansion evaluated today, we define an observational Hubble rate and deceleration parameter as
\ba
\left.H^\mathrm{obs}\right|_{0}\li=\li\left.K^\a K^\nu\del_\a u_\nu\right|_0,
\label{hubble def from series}\\
\left.q^\mathrm{obs}\right|_0\li=\li\left.\;{K^\a K^\nu K^\lambda\del_\a\del_\nu u_\lambda}
{(K^\kappa K^\e\del_\kappa u_\e)^2}\right|_0-3.\label{q0 from series}
\ea
We can also give an effective observed cosmological constant parameter from the $O(z^2)$ term:
\ba
\Omega_\Lambda^\mathrm{obs}&=&\frac52\left(1-q_0^\mathrm{obs}\right)-5+
\left.\frac{R_{\a \nu}K^\a K^\nu}{12(H^\mathrm{obs})^2}\right|_0
\nonumber\\&&
+\left.\frac{K^\a K^\nu K^\lambda K^\kappa\del_\a\del_\nu\del_\lambda u_\kappa}{6(H^\mathrm{obs})^3}\right|_0.
\ea
The argument of~\cite{HP} relies on proving that if all observers measure these three quantities to be isotropic then the spacetime is necessarily FLRW.
In a general spacetime~\cite{MacEllis}
\be
H_0^\mathrm{obs}={\frac{1}{3}\Theta}+
{A_\a e^\a}+
{\sigma_{\a\nu}e^\a e^\nu},
\label{hubble from series}
\ee
where ${A_\a e^\a}$ is a dipole and ${\sigma_{\a\nu}e^\a e^\nu}$ is a quadrupole.
Hence, if all observers measure $H_0^\mathrm{obs}$ to be isotropic, then $\sigma_{\a\nu}=0=A_\a$.
In a spacetime with isotropic $H_0^\mathrm{obs}$ the generalized deceleration parameter is given by~\cite{Cthesis}
\ba
\left(H_0^\mathrm{obs}\right)^2\, q_0^\mathrm{obs}&=&\;16\rho+\;12p-\;13\Lambda
 +\;{1}{5}e^\a\left[
2q_\a-{3}\sdel_\a\Theta
\right]
 \nonumber\\
&&+e^\a e^\nu\left[
 E_{\<\a\nu\>}-\;12\pi_{\<\a\nu\>}\right]
 .\label{q_0-observable}
\ea
If the dipole of this term vanishes then we see from Eq.~(\ref{c2}) that the energy flux must vanish as well as spatial gradients of the expansion. Excluding models with unphysical anisotropic pressure, Eq.~(\ref{e6}) then shows that the electric Weyl tensor must vanish, and it follows that the spacetime must be FLRW. The more general proof in~\cite{HP} uses $\Omega_\Lambda^\mathrm{obs}$ to show that the vorticity must necessarily vanish along with the anisotropic pressure.

It is an open but important question how this result translates to the case of almost-isotropy.

\section{COSMOLOGY WITHOUT THE COPERNICAN PRINCIPLE}

While the Copernican Principle remains untested, inhomogeneous models should be explored. As we discussed in the previous section isotropic observations imply spherical symmetry in the presence of dust matter and $\Lambda$, leading to the LTB model. A more general spherically symmetric perfect fluid solution is the Lemaitre model~\cite{AH}, which can include inhomogeneous radiation, or other types of barotropic perfect fluid. LTB solutions are the dust subcase.
An interesting  explanation for the dark energy problem in cosmology is one where  the underlying geometry of the universe is significantly inhomogeneous on Hubble scales. Spacetimes used in this context are usually LTB models~-- so-called void models, first introduced in~\cite{MT1}. These models can look like dark energy because we have direct access only to data on our lightcone and so we cannot disentangle temporal evolution in the scale factor from radial variations. Void models are considered ungainly compared with standard cosmology because naively they revoke the Copernican Principle, placing us at or very near the centre of the universe. It may be that the models used are very simplistic descriptions of inhomogeneity, and more elaborate inhomogeneous ones will conform to some version of the Copernican Principle and still satisfy observational constraints on isotropy.

Instead, we may think of LTB void models as smoothing all observables over the sky, thereby compressing all inhomogeneities into one or two radial degrees of freedom centred about us~-- and then we avoid placing ourselves `at the centre of the universe' in the standard way. In this sense, these models are a natural first step in developing a fully inhomogeneous description of the universe.  However, given the results of the last section it is not clear if this interpretation will work. Indeed, if we can extend the exact result that matter lightcone-isotopy $\rightarrow$ FLRW, to the almost-isotropic case, this could rule out radially inhomogeneous models with this interpretation. Furthermore, it has not been shown that a sky-averaged inhomogeneous model would give an LTB dust solution. It is likely that this is not the case because averaging and evolution do not commute.

Alternatively, within the multiverse context, one can imagine a vast universe in which our observable patch happens to be rather inhomogeneous.
We can even imagine a multiverse in which there are many void-like regions; even if we happened to be near the centre of one with a Hubble-scale inhomogeneity, this may be natural within a larger context -- perhaps in the same way we discovered that the Milky Way is not particularly special once understood in the context of a plethora of galaxies~\cite{uzan}. With this idea, we need not violate the Copernican Principle if we live near the centre of a Hubble scale void; rather, we should just change our perspective, and think of the Hubble scale as very small rather than very large.

Indeed, as argued in~\cite{uzan}, the anthropic `explanation' for the current value of $\Lambda$, which relies on a multiverse of some sort for its philosophical underpinning, necessitates the violation of the Copernican Principle -- simply because the vast majority of universe patches are nothing like ours, and not at all suitable for complex life.

\subsection{{\bf Hubble-scale voids}}

An inhomogeneous void  may be modelled as a spherically symmetric LTB model with metric
\ba
\label{LTBmetric2}
ds^2 = -dt^2 + \frac{\apar^2(t,r)}{1-\kapr}dr^2 + \aperp^2(t,r)r^2d\Omega^2\,,
\ea
where the radial ($\apar$) and angular ($\aperp$) scale factors are related by
$\apar = \p(\aperp r)/\p r$. The curvature $\kappa=\kappa(r)$ is not constant but is instead a free function.
The two scale factors define two Hubble rates:
\ba\label{H}
\Hperp= \Hperp(t,r) \equiv \frac{\dot a_\perp}{\aperp},~~~~\Hpar=\Hpar(t,r) \equiv \frac{\dot a_{\|}}{\apar},
\ea
and the Friedmann equation takes on its familiar form:
\ba
\frac{\Hperp^2}{\Hperpo^2}=\omegmo\aperp^{-3} + \omegko \aperp^{-2},
\ea
where $\omegmo(r)+ \omegko(r)=1$ and $\omegmo(r)$ is a free function, specifying the matter density parameter today. In general, $\Hperpo(r)$ is also free, but removing the decaying mode by enforcing a uniform bang time~\cite{silk,zibin} fixes this in terms of $\omegmo(r)$.

With one free function we can design models that give any distance modulus we like. If we choose $\omegmo(r)$ to reproduce exactly a LCDM $D(z)$, then the LTB model is a void with steep radial density profile which is, strictly speaking, non-differentiable at the origin if we want $q_0<0$~\cite{VFW} (see also~\cite{CBKH}). Much has been made of this non-differentiability, but it is irrelevant for this cosmological modelling. In essence, the fact that a sharp-profiled void can reproduce the LCDM $D(z)$ exactly is of mathematical interest only; a void model should be constrained directly by data, and not matched to a best fit LCDM model. If this is done, a smooth void is perfectly acceptable with present data~\cite{FLSC}, but this provides a feasible route to distinguish between a smooth void and LCDM~\cite{CFL}. However, it is impossible to tell the difference between an evolving dark energy FLRW model and a void model, using distance data alone.

\subsection{{\bf Perturbation theory in void models}}

An important open problem in inhomogeneous models is the modelling of structure formation. This is important partly because it provides a means for distinguishing between FLRW and LTB. One example of where we might see an effect is in the peak in the two-point matter correlation function attributed to the Baryon Accoustic Oscillations (BAO). It has been shown that if LTB perturbations evolve as in FLRW, then BAO can be decisive in ruling out voids~\cite{ZMS,gbh1}. Whether this assumption is valid however requires a full analysis of perturbations.

There have been two approaches so far:
\begin{enumerate}
\item Using a covariant 1+1+2 formalism which was developed for gauge-invariant perturbations of spherically symmetric spacetimes~\cite{1+1+2,CB2}. The full master equations for LTB have not yet been derived, but some progress has been made in the `silent' approximation, neglecting the magnetic part of the Weyl tensor~\cite{zibin,peter}.
\item Using a 2+2 covariant formalism~\cite{GS,GMG}, developed for stellar and black hole physics. The full master equations for LTB perturbations were presented in~\cite{CCF} (see also~\cite{Tomita:1997pt}).
\end{enumerate}

In FLRW cosmology, perturbations split into scalar, vector and
tensor modes that decouple from each other, and so evolve
independently (to first order). Such a split cannot usefully be performed in the same way in a spherically symmetric spacetime, as the
background is no longer spatially homogeneous, and modes written in this way couple together.  Instead, there exists a decoupling of the perturbations into two independent sectors, called `polar' (or even) and `axial' (or odd), which are analogous, but not equivalent, to scalar and vector modes in FLRW. These are based on how the perturbations transform on the sphere. Roughly speaking, polar modes are `curl' free on $S^2$ while axial modes are divergence free. Further decomposition may be made into spherical harmonics, so all variables are for a given spherical harmonic index $\ell$, and modes decouple for each $\ell$~-- analogously to $k$-modes evolving independently on an FLRW background. A full set of gauge-invariant variables were given by~\cite{GMG} who showed that there exists a natural gauge -- the Regge-Wheeler gauge -- in which all perturbation variables are gauge-invariant (rather like the longitudinal gauge in FLRW perturbation theory). Unfortunately, the interpretation of the gauge-invariant variables is not straightforward in a cosmological setting.

Most of the interesting physics happens in the polar sector, so we will discuss that case, following~\cite{CCF}. The general form of polar perturbations of the metric can be written, in Regge-Wheeler gauge, as
\ba
\mathrm{d} s^2 = &-\left[1+(2 \gmgeta-\gmgchi-\gmgk) Y \right] \mathrm{d} t^2 -\frac{2
a_{\parallel} \gmgpsi Y}{\sqrt{1- \kappa r^2}} \mathrm{d} t \mathrm{d} r \nonumber
\\
&+\left[1+(\gmgchi+\gmgk) Y \right] \frac{a_{\parallel}^2
\mathrm{d} r^2}{(1-\kappa r^2)} +a_{\perp}^2 r^2 (1+ \gmgk Y) \mathrm{d}\Omega^2, \label{gpolar}
\ea
where $\gmgeta(t,r)$, $\gmgchi(t,r)$, $\gmgk(t,r)$ and $\gmgpsi(t,r)$ are gauge-invariant variables. The notation here is such that a variable times the spherical harmonic $Y$ has a sum over $\ell,m$, e.g., $\varphi Y=\sum_{\ell=0}^\infty\sum_{m=-\ell}^{m=+\ell}\varphi_{\ell m}(x^A) Y_{\ell m}(x^a)$, where $x^a$ are coordinates on $S^2$, and $x^A=(t,r)$.  The general form of polar matter perturbations in this gauge is given by
\ba
\label{upolar}
u_{\mu} &=& \left[\hat{u}_A+ \left( \gmggamma \hat{n}_A +\frac{1}{2} h_{AB}
\hat{u}^B \right) Y, \gmgalpha Y_{:a} \right]\\
\label{rhopolar}
\rho &=& \rho^{LTB} (1+\gmgomega Y),
\ea
where $\gmgalpha$, $\gmggamma$ and $\gmgomega$ are gauge-invariant velocity and density perturbations and $h_{AB}$ is the metric perturbation in the $x^A$ part of the metric; a colon denotes covariant differentiation on the 2-sphere.  The unit vectors in the time and radial directions are
\be
\hat{u}^A = (1,0)\,,~~~
\hat{n}^A = \left( 0, \frac{\sqrt{1- \kappa r^2}}{a_{\parallel}} \right).
\ee

The elegance of the Regge-Wheeler gauge is that the gauge-invariant metric perturbations are master variables for the problem, and obey a coupled system of PDEs which are decoupled from the matter perturbations. The matter perturbation variables are then determined by the solution to this system. We outline what this system looks like for $\ell\geq2$; in this case $\eta=0$. The generalized equation for the gravitational potential is~\cite{CCF}:
\be
\ddot\varphi+4H_\perp\dot\varphi-2\frac{\kappa} {a_\perp^2}\varphi=S_\varphi(\chi,\varsigma).
\ee
The left hand side of this equation has exactly the form of the usual equation for a curved FLRW model, except that here the curvature, scale factor and Hubble rate depend on $r$. On the right, $S_\varphi$ is a source term which couples this potential to gravitational waves, $\chi$, and generalized vector modes, $\varsigma$. These latter modes in turn are sourced by $\varphi$:
\ba
&&-\ddot{\gmgchi} + \gmgchi^{\prime \prime} -3 H_{\parallel}\dot{\gmgchi} -2 W \gmgchi^{\prime} + \Bigg[ 16 \pi G \rho -\frac{6M}{a_{\perp}^3}
\nonumber\\&& ~~~~~~~~ -4 H_{\perp} (H_{\parallel}-H_{\perp})
-\frac{(\ell -1)(\ell +2)}{a_{\perp}^2 r^2}\Bigg] \gmgchi
=S_\chi(\gmgpsi,\gmgk),\\
&&\dot{\gmgpsi} + 2 H_{\parallel} \gmgpsi = -\gmgchi^{\prime}.
\ea
The prime is a radial derivative defined by
\be
X^{\prime} \equiv \frac{\sqrt{1-\kappa r^2}}{a_{\parallel}} X_{,r}.
\ee

The gravitational field is inherently dynamic even at the linear level, which is not the case for a dust FLRW model with only scalar perturbations. Structure may grow more slowly due to the dissipation of potential energy into gravitational radiation.
Since $H_\perp=H_\perp(t,r)$, $a_\perp=a_\perp(t,r)$ and $\kappa=\kappa(r)$, perturbations in each shell about the centre grow at different rates, and it is because of this that the perturbations generate gravitational waves and vector modes. This leads to a very complicated set of coupled PDEs to solve for each harmonic $\ell$.

In fact, things are even more complicated than they first seem. Since the scalar-vector-tensor decomposition does not exist in non-FLRW models, the interpretation of the gauge-invariant LTB perturbation variables is subtle. For example, when we take the FLRW limit we find that
\ba
\gmgk&=& -2\Psi-2\H V -2\frac{(1-\kappa r^2)}{r}h_r +\frac{1}{r^2}h^\T\nonumber\\&&
+\left[-\H \p_\tau  +\frac{(1-\kappa r^2)}{r} \p_r
 +\frac{\ell(\ell+1)-4(1-\kappa r^2)}{2r^2}\right]h^\TF,
\ea
where $\Psi$ is the usual perturbation space potential, $V$ is the radial part of the vector perturbation, and the $h$'s are invariant parts of the tensor part of the metric perturbation. Thus $\varphi$ contains scalars, vectors and tensors. A similar expression for $\varsigma$ shows that it contains both vector and tensor degrees of freedom, while $\chi$ is a genuine gravitational wave mode, as may be seen from the characteristics of the equation it obeys. This mode mixing may be further seen in the gauge-invariant density perturbation which appears naturally in the formalism:
\ba
8 \pi G \rho \gmgomega&=& -\gmgk^{\prime \prime} - 2 W \gmgk^{\prime}+(H_{\parallel}+2 H_{\perp}) \dot{\gmgk}+W
\gmgchi^{\prime} + H_{\perp} \dot{\gmgchi} \nonumber\\ &&+ \left[ \frac{\ell (\ell +1)}{a_{\perp}^2r^2} +2 H_{\perp}^2 +4 H_{\parallel} H_{\perp} -8 \pi G \rho \right]
(\gmgchi +\gmgk)  \nonumber\\ && - \frac{(\ell -1)(\ell +2)}{2 a_{\perp}^2r^2} \gmgchi  +2 H_{\perp} \gmgpsi^{\prime}+2 (H_{\parallel}+H_{\perp}) W
\gmgpsi ,
\ea
where
\be
W \equiv \frac{\sqrt{1-\kappa r^2}}{a_{\perp} r}.
\ee
When evaluated in the FLRW limit the mode mixing becomes more obvious still: $\Delta$ contains both vector and tensor modes, while its scalar part is
\be
4\pi G a^2\rho\,\gmgomega = \sdel^2\Psi-3\H\p_\tau\Psi-3(\H^2-\kappa)\Psi,
\ee
which gives the usual gauge invariant density fluctuation
$\delta\rho^\GI\equiv\delta\rho+\p_\tau\rho (B-\p_\tau
E)$~\cite{Malik&Wands}. Here,
$\sdel^2$ refers to the Laplacian
acting on a 3-scalar, so that
\be
\sdel^2=(1-\kappa r^2)\p_r^2+\frac{(2-3\kappa
  r^2)}{r}\p_r-\frac{\ell(\ell+1)}{r^2},
\ee
(which also helps one identify a Fourier $k$-mode in this context).
 The fact that $\Delta$ is more complicated is because the gauge-invariant density perturbation includes metric degrees of freedom in its definition; gauge-invariant variables which are natural for spherical symmetry may not be natural for homogeneous backgrounds. A gauge-dependent $\Delta$ may be defined which reduces to $\delta\rho^\GI$ in the FLRW subcase, but its gauge-dependence will cause problems in the inhomogeneous case.

A gauge-invariant variable which reduces to a pure scalar in the FLRW limit is~\cite{CCF}
\ba
\hat{\zeta}\equiv {\hat\lambda}''+2W{\hat\lambda}'- \frac{\ell(\ell+1)}{a_\perp^2
 r^2}{\hat\lambda}
+rW\hat\xi'+r \left(3W^2-\frac{1}{a_\perp^2r^2}\right)\hat\xi, \label{sc2}
\ea
where
\be
\hat{\lambda} \label{sc1} \equiv 8\pi G \rho a_\perp \left[H_\perp^{-1}\gmgomega-3\gmgalpha\right].
\ee
In the FLRW limit, $\hat\lambda$ contains no tensors. Furthermore,
\ba
\label{ve1}
\hat{\xi}&\equiv& \frac{3a_\perp}{2W}\Bigg[\frac{1}{r^3} \left(r^2\dot\gmgchi\right)'+\left(\frac{\gmgpsi}{r}\right)''+ 2W\left(\frac{\gmgpsi}{r}\right)'
\nonumber\\  &&~~~
-\left(\frac{\ell(\ell+1)-3}{a_\perp^2r^2}+3W^2\right) \frac{\gmgpsi}{r}\Bigg],\label{ve2}
\ea
is a pure vector in the FLRW limit.
In the FLRW limit, we find
\be
\zeta=\frac{8\pi G \rho}{\H}
\sdel^2\left[\frac{\delta\rho^\GI}{\rho}+ 3\mathcal{R}-3\Psi\right],
\ee
where the
curvature perturbation is
$\mathcal{R}=\Psi-\H(\H\Phi+\p_\tau\Psi)/ (\p_\tau\H-\H^2-\kappa)$~\cite{Malik&Wands}. These generalized scalar-vector-tensor variables may be useful in interpreting perturbations of void models.

These equations have not yet been solved in full generality. We expect different structure growth in LTB models, but it is not clear what form the differences will take. It seems reasonable to expect that the coupling between scalars, vectors and tensors will lead to dissipation in the growth of large-scale structure where the curvature gradient is largest, as it is the curvature and density gradients that lead to mode coupling. In trying to use structure formation to compare FLRW to LTB models, some care must be taken over the primordial power spectrum and whatever early universe model is used to generate perturbations -- since there is a degeneracy with the primordial power spectrum and the features in the matter power spectrum.

\subsection{{\bf The small-scale CMB in void models}}

The physics of decoupling and line-of-sight effects contribute differently to the CMB, and have different dependency on the cosmological model. In sophisticated inhomogeneous models both pre- and post-decoupling effects will play a role, but Hubble-scale void models allow an important simplification for calculating the moderate to high $\ell$ part of the CMB.

The comoving scale of the voids which closely mimic the LCDM distance modulus are typically $O(\mbox{Gpc})$. The physical size of the sound horizon, which sets the largest scale seen in the pre-decoupling part of the power spectrum, is around $150\,$Mpc.
This implies that in any causally connected patch of the Universe prior to decoupling, the density gradient is very small. Furthermore, the comoving radius of decoupling is larger than $10\,$Gpc, on which scale the gradient of the void profile is small anyway (by assumption). For example, at decoupling the total fractional difference in energy density  between the centre of the void and the asymptotic region is around 10\%~\cite{RC}; hence, across a causal patch we expect a maximum 1\% change in the energy density in the radial direction, and much less at the radius of the CMB that we observe for a Gaussian profile. This suggests that before decoupling on small scales we can model the universe in disconnected FLRW shells at different radii, with the shell of interest located at the distance where we see the CMB. This may be calculated using standard FLRW codes, but with the line-of-sight parts corrected for~\cite{ZMS,CFZ}.

For line-of-sight effects, we need to use the full void model. These come in two forms. The simplest effect is via the background dynamics, which affects the area distance to the CMB, somewhat similar to a simple dark energy model. This is the important effect for the small-scale CMB. The more complicated effect is on the largest scales through the Integrated Sachs-Wolfe effect (see~\cite{Tomita:2009yx} for the general formulas in LTB).  This requires the solution of the perturbation equations presented above.

\begin{figure}[t]
\begin{center}
\includegraphics[width=1.0\columnwidth]{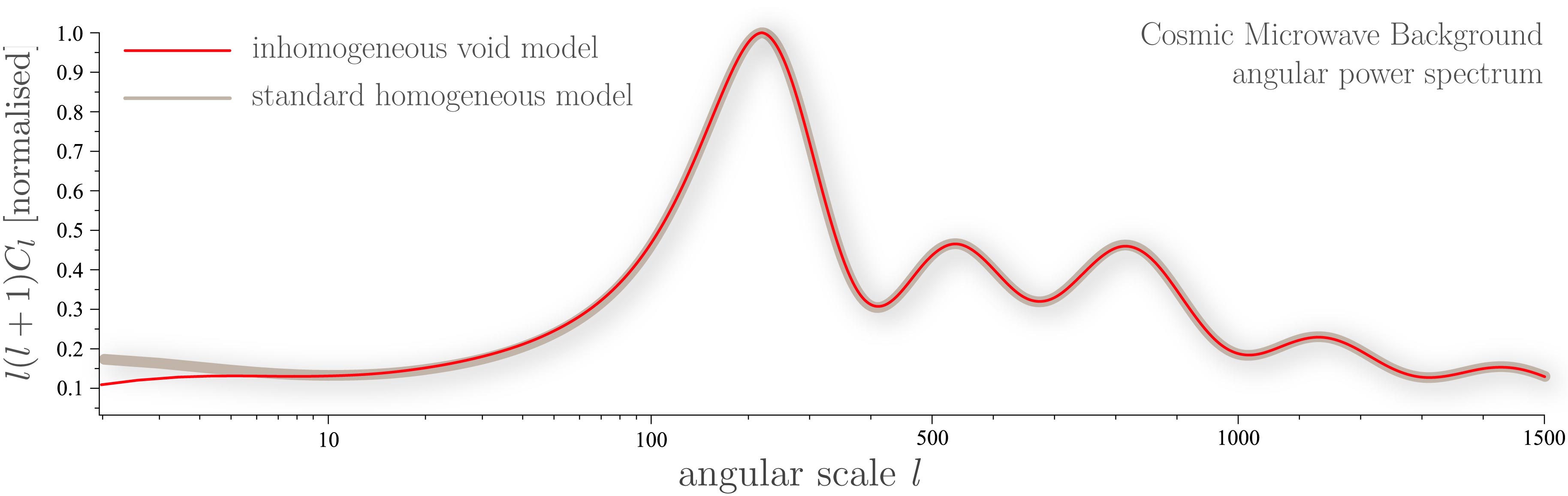}
\caption{{The normalized CMB angular power spectrum, from~\cite{RC}.} The power spectrum is shown against a default flat concordance model with zero tilt. There is nothing between the two models for high $\ell$, with the maximum difference around 1\%.}
\label{cmb}
\end{center}
\end{figure}

The CMB can be fit in void models in different ways. In~\cite{ZMS,CFZ}, it was shown that the CMB can be very restrictive on void models, although~\cite{CFZ} showed that with a varying bang time, the data for $H_0$, SNIa and CMB can be simultaneously accommodated.  Including inhomogeneous radiation in the background, it was argued in~\cite{RC} that the CMB can be accommodated along with other local observations with a homogeneous bang time, but with asymptotic curvature at the CMB radius. It is an open question exactly what constraints the small-scale CMB places on a generic void solution.
An example from~\cite{RC} of how closely a void model can reproduce the CMB power spectrum found in a concordance model is shown in Fig.~\ref{cmb}.
It is remarkable that the void models can reproduce the LCDM CMB power spectrum so closely. As emphasised in~\cite{CFZ}, it is often taken for granted that the CMB tells us that the universe is close to flat; these examples show that curvature can in fact be very large, but just not homogeneous.

\subsection{{\bf Summary of inhomogeneous models}}

An inhomogeneous LTB void model, even if it over-simplifies nonlinear inhomogeneity, does produce some rather remarkable results. The apparent acceleration of the universe can  be accounted for without dark energy.  It has also recently been shown~\cite{RC} that LTB models can simultaneously solve the so-called lithium problem, the apparent mismatch between $^7$Li abundance measured locally (thereby probing our past worldline at nucleosynthesis), and that inferred from measurements of deuterium and the baryon-photon ratio at the CMB (which probe nucleosynthesis at co-moving scales $O(\mbox{Gpc})$ away from us). In the concordance model, the lithium problem is a  $2.4-5.3\sigma$ discrepancy~\cite{Olive}, yet it can easily be accommodated with Hubble scale inhomogeneity in the baryon-photon ratio at early times.

It has also been shown that the small-scale CMB need not over-constrain voids and so rule them out~\cite{CFZ,RC}, but it seems that the simplest possibility~-- homogeneous bang time and asymptotic flatness~-- may not be possible.

The major open problems in inhomogeneous solutions to the dark energy problem  are:
\begin{itemize}
\item {\em CMB.}  The details of an inhomogeneous radiation era still have to be investigated. To calculate the large scale CMB, there are several additional effects which must be taken into account. The most important is the Sach-Wolfe effect, which requires knowing the perturbation spectrum at the time of decoupling on the largest angular scales. While it might not deviate from FLRW for a central observer because of the reasons discussed above, this needs further investigation.  In addition, the Integrated Sachs-Wolfe effect will contribute to the line-of-sight part of the CMB calculation~\cite{Tomita:2009yx,Tomita:2009wz}.

\item {\em Structure formation.} The complexity of the perturbation equations is a major stumbling block. Attempts have been made to solve a limited subcase of the equations~\cite{zibin,CCF,peter}, but it is not clear how close those approximations are to the full solution. This is required for a reliable estimate of corrections to the power spectrum and to the BAO peak.
\item {\em Inflation.} How can a void model arise in the first place? What kind of voids are possible from an inflationary era? Is there a natural way to produce many voids, and so ease the Copernican problem? Standard inflation can't easily give a single void, but one can imagine an inflationary potential~-- or perhaps a separate earlier period of inflation some 50 efolds before the standard slowly rolling inflationary phase~-- which imprints an extra scale in the primordial power spectrum.

\item {\em The Copernican problem.} As we have discussed above, it is not easy to find an inhomogeneous model with nearly isotropic observations which satisfies the Copernican Principle. Further work is needed to see if it is possible to avoid or understand the very unsatisfactory fine-tuning problem which exists in LTB voids: that we are within tens of Mpc of the centre~\cite{Alnes:2006pf,alnes,mortsell}. One way one can imagine this question being addressed is some kind of anthropic selection effect which makes intelligent life more likely in regions where the dark matter concentration is lowest.\footnote{S. R\"as\"anen, private communication.} For example, one could imagine that dark matter inhibits the formation of stable solar systems of sufficient longevity. Therefore, a naive calculation of a 1 in $\sim10^9$ chance of us being within tens of Mpc of the centre of a Hubble-scale void would have to be weighted appropriately, thereby reducing this fine-tuning problem. 
\end{itemize}

\section{DISCUSSION}

Without a theoretical understanding of the value of the cosmological constant the concordance model remains phenomenological no matter how strongly observations appear to support the model. The concordance model suffers an important coincidence problem: that dark energy starts to dominate at around the time our solar system forms.
Until we demonstrate observationally that the Universe is homogeneous on large scales, we should consider inhomogeneous spacetimes even if they are philosophically uncomfortable, particularly in light of the fact that in their simplest incarnation they can explain away the dark energy problem through inhomogeneity, without apparently causing other problems. It is worth reflecting on how remarkable this is. Measurements of a nonzero $\Lambda$ may in fact be thought of as an explicit consequence of the Copernican assumption.

It is important therefore to re-examine the basic assumptions of the standard concordance model -- both to test the Copernican Principle at the heart of modern cosmology, and to understand what freedom we have to develop new cosmological models that may be able to explain cosmic acceleration using curvature and inhomogeneity rather than a cosmological constant. In order to do this, we need non-perturbative methods and also new types of nonlinear perturbations where the nonlinearity arises from the inhomogeneous background. Moreover, the whole inflationary paradigm needs to be re-examined to see if inhomogeneities this large could arise naturally.

We have shown how the Copernican assumption when combined with the high isotropy of the CMB implies FLRW under fairly weak assumptions~-- mainly, that the CMB rest frame is geodesic. Provided that the dark matter is CDM that is comoving with the CMB, this geodesic property will hold. But in interacting dark energy-dark matter models, for example, the geodesic radiation frame is an additional assumption.  In order to test the Copernican Principle at the foundation of the FLRW models, observations which look inside our past lightcone to estimate how distant observers see the CMB~\cite{gbh2,goodman,CS,Jia:2008ti} are necessary. In order to observationally prove FLRW, however, we also must demonstrate that galaxies follow geodesics, and that dark energy is comoving with the CMB.

If galaxies were non-geodesic, then this would leave a signature as a dipole in the Hubble law which grows linearly with distance (and so is distinguishable from a normal bulk flow, which is constant)~\cite{CB}
which is derived from the term which is linear in redshift in the area distance-redshift relation~-- see Eq.~(\ref{hubble from series}).
Possible dark energy flux would leave a similar signal in the quadratic term in the area distance-redshift relation. The generalized deceleration parameter, in a fully inhomogeneous spacetime is, following~\cite{Cthesis},
\ba
H_0^2\, q_0&=&\;16\rho+\;12p-\;13\Lambda
 -\;23\D_\a\udot^\a
 +\;{12}{5}\sigma_{\a\nu} \sigma^{\a\nu}-\;23\omega_\a \omega^\a
  \nonumber\\&&
 +e^\a\Bigg[-\;23\Theta \udot_\a
-\;{3}{5}\sdel_\a\Theta-\dot
 A_\a+\udot^\nu\sigma_{\a\nu}+\;25q_\a
 \nonumber\\ &&
 ~~~~~~~~~+\;35\varepsilon_\a^{~ \nu\gamma}\left(\sdel_\nu\omega_\gamma
 +2\udot_\nu\omega_\gamma\right)\Bigg]
 \nonumber\\
&& \left.+e^\a e^\nu\left[-2\sdel_{\<\a}\udot_{\nu\>}+\udot_{\<\a}\udot_{\nu\>}
 +E_{\<\a\nu\>}-\;12\pi_{\<\a\nu\>}+
\;97\sigma^{\vphantom{c}}_{\gamma\<\a} \sigma_{\nu\>}^{\phantom{\nu\>}\gamma}
 \right.\right. \nonumber\\
 &&\left.\left.
 +\omega_{\<\a}\omega_{\nu\>}-
\;23\sigma^{\vphantom{c}}_{\gamma\<\a} \omega_{\nu\>}^{\phantom{\nu\>}\gamma}
 \right]\right. +
e^\a e^\nu e^\gamma\left[-5\udot_{\<\a}\sigma_{\nu\gamma\>}-\sdel_{\<\a} \sigma_{\nu\gamma\>}\right]\nonumber\\
&&
 -3e^\a e^\nu e^\gamma e^\delta\sigma_{\<\a\nu}\sigma_{\gamma\delta\>}
 .\label{q_0-observable}
\ea
If it were confirmed observationally that the CMB frame is geodesic, and that the CMB is almost isotropic around distant observers, then this $H_0^2q_0$ should be an isotropic observable~-- a potentially useful consistency check.

Although it seems obvious that CDM and baryons are just geodesic dust, we have to be careful because of issues highlighted from the averaging problem. While a distribution of non-relativistic CDM particles obey a dust equation of state at all times, more and more of the CDM particles are locked up as galaxies and clusters form, and yet those virialized clusters then become the `dust particles', within the standard paradigm. Are these geodesic? It is not yet clear how to treat this situation properly, and it is conceivable that there is a backreaction effect which changes the effective equation of state of matter at late times~\cite{BR} (amongst other effects).
Depending on precisely what is calculated, this backreaction can be divergent in the concordance model~\cite{CAL}, and remains a significant problem to calculate non-linearly (although see~\cite{Baumann:2010tm} who find no such divergences using a different method of averaging).

Off-lightcone observations will be able to place limits on the spatial gradients of the CMB multipoles as well as the multipoles themselves, which we have shown is also a requirement for deducing almost-FLRW geometry from almost-isotropic CMB (the almost-EGS result). The important exact ETM result states that if the first three multipoles are zero then all the rest must be too. It seems reasonable that this would extend to an almost-ETM result if the first 3 multipoles are small. In this case observations which look inside our past lightcone might only need to confirm that the first three multipoles around distant observers are small in order to observationally confirm FLRW. These issues deserve further investigation, and are important because it might be possible to construct a genuinely inhomogeneous model which solves the dark energy problem, but which also satisfies the Copernican principle, maintaining the isotropy of the CMB for all or `most' observers (as claimed in~\cite{Wiltshire:2007zj}, for example).


Dropping the Copernican Principle, we have the key exact result that isotropy of matter observations down our past lightcone implies an LTB geometry. However, it is an open question whether almost-isotropy of matter observations on the past lightcone implies an almost-spherically symmetric LTB model. We have discussed a complete analysis of structure growth at first order, which is the next step in developing LTB models, and shown how this is much more involved than the linear perturbations of FLRW. However, we argue that it is worth the effort. All tests of the Copernican Principle, including those which can be done by examining the background dynamics down our past lightcone~\cite{CBL,UCE}, are difficult and sensitive. Only by developing a family of inhomogeneous spacetimes to the same level of sophistication as the standard concordance model, and directly comparing them side by side, will we really be able to understand whether $\Lambda$ is real, or actually a consequence of our homogeneity assumption.

~\\{\bf Acknowledgments:}
We thank Timothy Clifton, Ruth Durrer,  George Ellis, Pedro Ferreira, David Matravers, John Moffat, Syksy R\"as\"anen and Marco Regis for comments and/or discussions. CC is supported by the National Research Foundation (NRF, South Africa) and RM is supported by the UK Science \& Technology Facilities Council. This work was partially supported by the South African National Institute for Theoretical Physics and by a Royal Society (UK)/ NRF exchange grant between the Universities of Portsmouth and Cape Town.

\newpage
\appendix

\section{Fully nonlinear field and Boltzmann equations}

We summarize, using~\cite{MGE}, the key equations in the 1+3 covariant Ehlers-Ellis formalism (see also~\cite{Langlois:2010vx} in this volume, and~\cite{EvE,TCM} for reviews). This formalism provides a physically transparent formulation of the field equations and the Boltzmann equation in full nonlinear generality. The formalism leads to a simple `top-down' approach to linear perturbations around FLRW -- i.e., one starts from the nonlinear spacetime and then takes the FLRW limit. By contrast, the `bottom-up' approach of the usual coordinate-based perturbation theory (see~\cite{Malik&Wands} for a review), starts from FLRW and perturbs away from it. In the context of perturbed FLRW models, the covariant approach has no advantage over the coordinate-based approach. But beyond the FLRW framework, when one cannot assume there is an FLRW background to perturb away from, the coordinate-based approach cannot be applied, while the covariant approach is applicable.

The Ehlers-Ellis formalism is a covariant Lagrangian approach to gravitational dynamics, based on a decomposition relative to a chosen 4-velocity field $u^\a$.
The fundamental tensors are
\begin{equation}\label{hep}
h_{\a\nu}=g_{\a\nu}+u_\a u_\nu,~ ~ \ep_{\a\nu\alpha}=\eta_{\a\nu\alpha\beta}u^\beta,
\end{equation}
where $h_{\a\nu}$ projects into the instantaneous rest space of comoving
observers, and $\ep_{\a\nu\alpha}$ is the projection of the spacetime alternating tensor $\eta_{\a\nu\alpha \beta}=-\sqrt{-g}
\delta^0{}_{[\a}\delta^1{}_\nu\delta^2{}_\alpha\delta^3{}_{\beta]}$.
The projected symmetric tracefree (PSTF) parts of vectors and
rank-2 tensors are
\begin{eqnarray}
V_{\langle \a\rangle}=h_\a{}^\nu V_\nu\,,~ S_{\langle \a\nu \rangle }=
\Big\{h_{(a}{}^\alpha h_{\nu)}{}^\beta-
{{1\over3}}h^{\alpha \beta }h_{\a\nu }\Big\}S_{\alpha \beta }\,. \label{pstf}
\end{eqnarray}
The skew part of a
projected rank-2 tensor is spatially dual to the projected vector,
$S_{\a}={1\over2}\ep_{\a\nu \alpha}S^{[\nu \alpha]}$, and then any projected rank-2
tensor has the decomposition
$S_{\a\nu }={1\over 3}Sh_{\a\nu }+\ep_{\a\nu \alpha}S^{\alpha}+S_{\langle \a\nu \rangle}$, where $S=S_{\alpha\beta} h^{\alpha \beta}$.

The covariant derivative $\nabla_{\a}$ defines 1+3 covariant time and spatial derivatives:
\begin{eqnarray}
\dot{J}^{\a\cdots}{}{}_{\cdots \nu}= u^{\alpha} \nabla_{\alpha}
J^{\a\cdots}{}{}_{\cdots \nu},\, \D_{\alpha} J^{\a\cdots}{}{}_{\cdots \nu} =    h_{\alpha}{}^\beta h^{\a}{}_\kappa\cdots h_{\nu}{}^\tau
\nabla_\beta J^{\kappa\cdots}{}{}_{\cdots \tau}. \label{dd}
\end{eqnarray}
The projected derivative $\D_{\a}$
defines a covariant PSTF divergence, $\D^\a V_\a\,, ~ \D^\nu S_{\a\nu}$,
and a covariant PSTF curl,
\begin{eqnarray}
\mbox{curl}\, V_{\a}=\ep_{\a\nu c}\D^{\nu}V^{\alpha}\,,~ \mbox{curl}\,
S_{\a\nu }=\ep_{\alpha \beta (\a}\D^{\alpha}S_{\nu)}{}^\beta\,. \label{curl}
\end{eqnarray}

The relative motion of comoving observers
is encoded in the PSTF kinematical quantities: the volume expansion rate,  4-acceleration, vorticity
and shear, given respectively by
\begin{eqnarray}
\Theta=\D^{\a}u_{\a},~ A_{\a}=\dot{u}_{\a},~ {\omega}_\a=\mbox{curl}\, u_\a ,~ \sigma_{\a\nu }=\D_{\langle \a}u_{\nu\rangle }. \label{kin}
\end{eqnarray}
Thus
\begin{eqnarray}
\nabla_{\nu}u_{\a}={{1\over3}}\Theta h_{\a\nu }+\ep_{\a\nu \alpha}\omega^{\alpha}
+\sigma_{\a\nu }-A_{\a}u_{\nu}\,. \label{du}
\end{eqnarray}
A key identity (valid in the fully nonlinear case) is
\begin{eqnarray}
\mbox{curl}\,\D_{\a}\psi \equiv \ep_{\a\nu \alpha}\D^{\nu}\D^{\alpha}\psi=
-2\dot{\psi}\omega_{\a} \,, \label{ri1}
\end{eqnarray}
which shows that curl grad is nonzero in the presence of vorticity (a purely relativistic feature, with no Newtonian analogue). A crucial nonlinear commutation relation for scalars is
\be
h_\a^{~\nu}(\D_{\nu}\psi\dot)-\D_{\a}\dot\psi = \dot\psi A_\a-\left(\frac{1}{3}\Theta h_{\a\nu}+\sigma_{\a\nu}+\varepsilon_{\a\nu \kappa}\omega^\kappa\right)\D^\nu\psi\, .\label{timespace}
\ee

The PSTF dynamical quantities describe the
sources of the gravitational field:
the (total) energy density $\rho=T_{\a\nu }u^{\a}u^{\nu}$, isotropic pressure
$p={1\over3}h_{\a\nu }T^{\a\nu }$, momentum density $q_{\a}=-T_{\langle \a\rangle \nu}u^{\nu}$,
and anisotropic stress $\pi_{\a\nu }=T_{\langle \a\nu \rangle}$, where $T_{\a\nu }$ is
the total energy-momentum tensor. The locally free gravitational
field, i.e. the part of the spacetime curvature not directly
determined locally by dynamical sources, is given by the Weyl tensor
$C_{\a\nu \alpha \beta }$. This splits into the PSTF gravito-electric and gravito-magnetic fields
\begin{eqnarray}
E_{\a\nu }=C_{\a \alpha \nu \beta}u^{\alpha}u^\beta\,,~~
H_{\a\nu }={{1\over2}}\ep_{\a\alpha \beta }C^{\alpha \beta }{}{}_{\nu \kappa}u^\kappa  \,,
\end{eqnarray}
which provide a covariant description of tidal forces
and gravitational radiation.

The Ricci and Bianchi identities,
\begin{eqnarray}
\nabla_{[\a} \nabla_{\nu]} u_\alpha= R_{\a\nu \alpha  \beta}u^{\beta}, ~~~ \nabla^\beta
C_{\a\nu \alpha \beta } =- \nabla_{[\a}\Big\{ R_{\nu]\alpha} - {1\over6}Rg_{\nu]\alpha} \Big\}, \label{rbi}
\end{eqnarray}
produce the
fundamental evolution and constraint equations governing the
covariant quantities. Einstein's equations are
incorporated via the algebraic replacement of the Ricci tensor
\begin{equation}\label{efe}
R^{\a\nu }=T^{\a\nu }-{1\over2}T_{\alpha}{}^{\alpha}g^{\a\nu },
\end{equation}
where $T^{\a\nu}$ is the total energy-momentum tensor, including a cosmological constant component, $T_\Lambda^{\a\nu}=-\Lambda g^{\a\nu}$.

The resulting equations, in fully nonlinear form and for a general
source of the gravitational field, are:

\noindent{\em Evolution:}
\begin{eqnarray}
&& \dot{\rho} +(\rho+p)\Theta+\D^\a q_\a = -2A^{\a} q_{\a}
 -\sigma^{\a\nu }\pi_{\a\nu }\,, 
 \label{e1}\\
&& \dot{\Theta} +{{1\over3}}\Theta^2 +{{1\over2}}(\rho+3p)-\D^\a
A_\a \nonumber\\
&&~~~~~~~ = -\sigma_{\a\nu }\sigma^{\a\nu }
+2\omega_{\a}\omega^{\a}+A_{\a}A^{\a} \,,
\label{e2}\\
&& \dot{q}_{\langle \a\rangle }
+{{4\over3}}\Theta q_{\a}+(\rho+p)A_{\a} +\D_{\a} p +\D^\nu\pi_{\a\nu} \nonumber\\
&&~~~~~~~ = -\sigma_{\a\nu }q^{\nu}
+\ep_{\a\nu\alpha}\omega^\nu q^{\alpha} -A^{\nu}\pi_{\a\nu } \,,
\label{e3} \\
&& \dot{\omega}_{\langle \a\rangle } +{{2\over3}}\Theta\omega_{\a}
+{{1\over2}}\mbox{curl}\, A_{\a} = \sigma_{\a\nu }\omega^{\nu} \,,\label{e4}\\
&& \dot{\sigma}_{\langle \a\nu \rangle } +{{2\over3}}\Theta\sigma_{\a\nu }
+E_{\a\nu }-{{1\over2}}\pi_{\a\nu } -\D_{\langle \a}A_{\nu\rangle } \nonumber\\
&&~~~~~~~ = -\sigma_{\alpha\langle \a}\sigma_{\nu\rangle }{}^{\alpha} - \omega_{\langle \a}\omega_{\nu\rangle }
+A_{\langle \a}A_{\nu\rangle }\,,
\label{e5}\\
&& \dot{E}_{\langle \a\nu \rangle } +\Theta E_{\a\nu }
-\mbox{curl}\, H_{\a\nu } +{{1\over2}}(\rho+p)\sigma_{\a\nu }\nonumber\\
&&~~~~~~~ +{{1\over2}}
\dot{\pi}_{\langle \a\nu \rangle } +{{1\over6}}
\Theta\pi_{\a\nu } +{{1\over2}}\D_{\langle \a}q_{\nu\rangle } \nonumber\\
&&~~~~~~~ =-A_{\langle \a}q_{\nu\rangle } +2A^{\alpha}\ep_{\alpha \beta (\a}H_{\nu)}{}^\beta
+3\sigma_{\alpha\langle \a}E_{\nu\rangle }{}^{\alpha} \nonumber\\
&&~~~~~~~~~  -\omega^{\alpha} \ep_{\alpha \beta (\a}E_{\nu)}{}^\beta -{{1\over2}}\sigma^{\alpha}{}_{\langle
\a}\pi_{\nu\rangle \alpha} -{{1\over2}}\omega^{\alpha}\ep_{\alpha \beta (\a}\pi_{\nu)}{}^\beta \,,
\label{e6}\\
&& \dot{H}_{\langle \a\nu \rangle } +\Theta H_{\a\nu } +\mbox{curl}\, E_{\a\nu }
-{{1\over2}}\mbox{curl}\,\pi_{\a\nu } \nonumber\\
&&~~~~~~~ = 3\sigma_{\alpha\langle \a}H_{\nu\rangle }{}^{\alpha}
-\omega^{\alpha} \ep_{\alpha \beta (\a}H_{\nu)}{}^\beta \nonumber\\
&&~~~~~~~~~ -2A^{\alpha}\ep_{\alpha \beta (\a}E_{\nu)}{}^\alpha -{{3\over2}}\omega_{\langle \a}q_{\nu\rangle
}+ {{1\over2}}\sigma^{\alpha}{}_{(\a}\ep_{\nu)\alpha \beta }q^\beta \,. \label{e7}
\end{eqnarray}

\noindent{\em Constraint:}
\begin{eqnarray}
&& \D^\a\omega_\a = A^{\a}\omega_{\a} \,, ~~~~~~~~~~~~~~~~~~~~~~~~~~~~~ ~~~~~~~~~~~~~~~~~~~~~~~~~~~~~~~~~~~~~~~~~ \label{c1}\\
&& \D^\nu\sigma_{\a\nu}-\mbox{curl}\,\omega_{\a} -{{2\over3}}\D_{\a}\Theta +q_{\a} =-
2\ep_{\a\nu\alpha}\omega^\nu A^{\alpha}  \,,\label{c2}\\
&&  \mbox{curl}\,\sigma_{\a\nu }+\D_{\langle \a}\omega_{\nu\rangle }
 -H_{\a\nu }= -2A_{\langle \a}
\omega_{\nu\rangle } \,,\label{c3}\\
&& \D^\nu E_{\a\nu}
+{{1\over2}}\D^\nu\pi_{\a\nu}
 -{{1\over3}}\D_{\a}\rho
+{{1\over3}}\Theta q_{\a} \nonumber\\
&& ~~~~~~= \ep_{\a\nu\alpha}\sigma^\nu{}_\beta H^{\alpha \beta} -3H_{\a\nu}
\omega^{\nu} +{{1\over2}}\sigma_{\a\nu }q^{\nu}-{{3\over2}}
\ep_{\a\nu\alpha}\omega^\nu q^{\alpha}  \,,\label{c4}\\
&& \D^\nu H_{\a\nu}
+{{1\over2}}\mbox{curl}\, q_{\a}
 -(\rho+p)\omega_{\a}\nonumber\\
&& ~~~~~~ =
-\ep_{\a\nu\alpha}\sigma^\nu{}_\beta E^{\alpha \beta}-{{1\over2}}\ep_{\a\nu\alpha}\sigma^\nu{}_\beta \pi^{\alpha \beta}   +3E_{\a\nu }\omega^{\nu}
-{{1\over2}}\pi_{\a\nu } \omega^{\nu}  \,.\label{c5}
\end{eqnarray}

The energy and momentum conservation equations are the evolution equations~(\ref{e1}) and (\ref{e3}). The dynamical quantities $\rho, p, q_\a, \pi_{\a\nu}$ in the evolution and constraint equations
(\ref{e1})--(\ref{c5}) are the total quantities, with
contributions from all dynamically significant particle species.
Thus
\begin{eqnarray}
T^{\a\nu } &=& \sum_I T_{I}^{\a\nu } = \rho
u^{\a}u^{\nu}+ph^{\a\nu }+2q^{(\a}u^{\nu)}
+\pi^{\a\nu } \,, \label{t1}\\
T_{I}^{\a\nu }&=&
\rho^*_{I}u_{I}^{\a}u_{I}^{\a}+p^*_{I}h_{I}^{\a\nu }
+2q_{I}^{*(\a}u_{I}^{\nu)}+\pi_{I}^{*\a\nu }\,, \label{t2}
\end{eqnarray}
where $I=r,n,b,c,\Lambda$ labels the species. The asterisk on the
dynamical quantities $\rho^*_{I},\cdots$ is intended to emphasize that these quantities are measured, not in the $u^\a$-frame, but in the $I$-frame, whose 4-velocity is given by
\begin{eqnarray}
u_{I}^{\a}=\gamma_{I}\left(u^{\a}+v_{I}^{\a}\right)\,, ~v_{I}^{\a}u_{\a}=0\,, ~\gamma_I=\left( 1-v_I^2 \right)^{-1/2}.\label{t3}
\end{eqnarray}
The fully nonlinear equations for the $I$ dynamical quantities as measured in the fundamental $u^{\a}$-frame are:
\begin{eqnarray}
\rho_{I} &=& \rho^*_{I} +
\Big\{\gamma_{I}^2v_{I}^2\left(\rho^*_{I}+p^*_{I}\right)
+2\gamma_{I}q_{I}^{*\a}
v_{I\a}+\pi_{I}^{*\a\nu }v_{I\a}v_{I\nu}\Big\} \,,\label{t4}\\
p_{I} &=&  p^*_{I} +{{1\over3}}
\Big\{\gamma_{I}^2v_{I}^2\left(\rho^*_{I}
+p^*_{I}\right)+2\gamma_{I}q_{I}^{*\a}
v_{I\a}+\pi_{I}^{*\a\nu }v_{I\a}v_{I\nu}\Big\}\,, \label{t5}\\
q_{I}^{\a} &=& q_{I}^{*\a}+(\rho^*_{I}+p^*_{I})v_{I}^{\a}
+\Big\{ (\gamma_{I}-1)q_{I}^{*\a}
-\gamma_{I}q_{I}^{*\nu}v_{I\nu}u^{\a}+ \gamma_I q^{*\nu}_I v_{I\nu} v_I^\mu\nonumber\\
&&{} +\gamma_{I}^2v_{I}^2
\left(\rho^*_{I}+p^*_{I}\right)v_{I}^{\a}
+\pi_{I}^{*\a\nu }v_{I\nu}-\pi_{I}^{*\nu\alpha} v_{I\nu}v_{I\alpha}u^{\a}
\Big\} \,,
\label{t6}\\ 
\pi_{I}^{\a\nu } &=& \pi_{I}^{*\a\nu } +
\Big\{-2u^{(\a}\pi_{I}^{*\nu)\alpha}v_{I\alpha}+\pi_{I}^{*\nu\alpha} v_{I\nu}
v_{I\alpha}u^{\a}u^{\nu} +2\gamma_{I}v_{I}^{\langle \a}q_{I} ^{*\nu\rangle} \nonumber\\
&&{}- 2 \gamma_I q^{*\alpha}_I v_{I\alpha} u^{(\mu} v_I^{\nu)} -{1\over 3}\pi_{I}^{*\alpha \beta }v_{I\alpha}v_{I\beta}h^{\a\nu } +
\gamma_{I}^2\left(\rho^*_{I}+p^*_{I}\right) v_{I}^{\langle
\a}v_{I}^{\nu\rangle}
\Big\}\,.\nonumber\\ \label{t7}
\end{eqnarray}
The terms in braces are the nonlinear corrections that vanish in the standard perturbed FLRW case.
The total dynamical quantities in Eqs.~(\ref{e1})--(\ref{c5}), are given by
\begin{eqnarray}
\rho=\sum_I\rho_{I}\,,~p=\sum_I p_{I}\,,~ q^{\a}=\sum_I
q_{I}^{\a}\,,~\pi^{\a\nu }=\sum_I\pi_{I}^{\a\nu }\,.
\end{eqnarray}
Assuming that the species are non-interacting, they each separately obey the energy and momentum conservation equations~(\ref{e1}) and (\ref{e3}):
\begin{eqnarray}
&& \dot{\rho}_I +(\rho_I+p_I)\Theta+\D_\a q_I^\a = -2A_{\a} q_I^{\a}
 -\sigma_{\a\nu }\pi_I^{\a\nu }\,, 
 \label{e1i}\\
&& \dot{q}_I^{\langle \a\rangle }
+{{4\over3}}\Theta q_I^{\a}+(\rho_I+p_I)A^{\a} +\D^{\a} p_I +\D_\nu\pi_I^{\a\nu} \nonumber\\
&&~~~~~~~ = -\sigma^\a{}_\nu q_I^{\nu}
+\ep^\a{}_{\nu\alpha}\omega^\nu q_I^{\alpha} -A_{\nu}\pi_I^{\a\nu } \,,
\label{e3i}
\end{eqnarray}
where the $I$-quantities are given by Eqs.~(\ref{t4})--(\ref{t7}).

The Ehlers-Ellis covariant kinetic theory description starts by splitting the photon 4-momentum as
\begin{equation}
p^{\a}=E(u^{\a}+e^{\a})\,,~~e^{\a} e_{\a}=1\,,~e^{\a} u_{\a}=0\,. \label{E}
\end{equation}
Here $E=-u_{\a}p^{\a}$ is the energy and $e^{\a}=p^{\langle \a\rangle}/E$ is the
direction, as measured by a comoving fundamental observer. Then
the photon distribution function is decomposed into covariant
harmonics via the expansion
\begin{eqnarray}
f(x,p)=f(x,E,e) &=& F+F_{\a}e^{\a}+F_{\a\nu }e^{\a}e^{\nu}+\cdots \nonumber \\ &=& \sum_{\ell\geq0}
F_{M_\ell}(x,E) e^{\langle M_\ell\rangle}, \label{r3}
\end{eqnarray}
where
${M_\ell}\equiv {\a_1}{\a_2}\cdots {\a_\ell}$ and $e^{M_\ell} \equiv e^{\a_1}\cdots e^{\a_\ell}$. The multipoles $F_{M_\ell}$ are a covariant alternative to the usual expansion in spherical harmonics. They are PSTF:
\begin{eqnarray}
F_{\a\cdots \nu}=F_{\langle \a\cdots \nu\rangle}\,\Leftrightarrow \, F_{\a\cdots
\nu}=F_{(\a\cdots \nu)},F_{\a\cdots \nu}u^{\nu}=0= F_{\a\cdots \alpha\beta}h^{\alpha\beta}.~~~~~~\label{r3'}
\end{eqnarray}

The first 3 multipoles determine the radiation energy-momentum
tensor,
\begin{eqnarray}
T_{r}^{\a\nu }(x)& \equiv &\int p^{\a}p^{\nu}f(x,p)\mathrm{d}^3p \nonumber\\ &=&
\rho_{r}u^{\a}u^{\nu}+{{1\over3}}\rho_{r}h^{\a\nu }
+2q_{r}^{(\a}u^{\nu)}+\pi_{r}^{\a\nu }\,,
\end{eqnarray}
where $\mathrm{d}^3p=E\mathrm{d} E\mathrm{d}\Omega$ is the covariant volume element on the
future light cone at event $x$. It follows that the dynamical quantities of the radiation (in the $u^{\a}$-frame) are: \begin{eqnarray}
\rho_{r} &=& 4\pi\int_0^\infty E^3F\,\mathrm{d} E\,, ~q_{r}^{\a} =
{4\pi\over 3}\int_0^\infty E^3F^{\a}\,\mathrm{d} E \,,\nonumber\\ \pi_{r}^{\a\nu } &=&
{8\pi\over 15}\int_0^\infty E^3F^{\a\nu }\,\mathrm{d} E\,. \label{em3} \end{eqnarray}
We extend these dynamical quantities to all multipole orders by
defining the brightness multipoles
\begin{eqnarray}
\Pi_{\a_1\cdots \a_\ell} = \int E^3
F_{\a_1\cdots \a_\ell}\mathrm{d} E\,, \label{r10}
\end{eqnarray}
so that
\begin{equation}\label{}
\Pi={1\over 4\pi}\rho_{r},~ \Pi^{\a}={3\over 4\pi}q_{r}^{\a}, ~\Pi^{\a\nu }={15\over 8\pi}\pi_{r}^{\a\nu }.
\end{equation}

The collisionless Boltzmann (or Liouville) equation is
\begin{equation}
{\mathrm{d} f\over \mathrm{d} v}\equiv p^{\a}{\p f\over \p x^{\a}}-\Gamma^{\a}{}_{\alpha\beta}
p^{\alpha}p^{\beta}{\p f\over \p p^{\a}}=0 \,, \label{boltz}
\end{equation}
where $p^{\a}=\mathrm{d} x^{\a}/\mathrm{d} v$, and we neglect polarization.
The covariant multipoles of $\mathrm{d} f/\mathrm{d} v$ are given by
\begin{eqnarray}
&&{1\over E}\left({\mathrm{d} f \over \mathrm{d} v}\right)_{M_\ell} =\dot{F}_{\langle M_\ell \rangle}-{{1\over3}}\Theta
EF'_{M_\ell} +\D_{\langle \a_\ell} F_{M_{\ell-1}\rangle}
+{(\ell+1)\over(2\ell+3)} \D^{\a}F_{\a M_\ell} ~~~~
\nonumber\\
&&{}
-{(\ell+1)\over(2\ell+3)}E^{-(\ell+1)}\left[E^{\ell+2}F_{\a M_\ell}
\right]'A^{\a}-E^\ell\left[E^{1-\ell}F_{\langle M_{\ell-1}}\right]'
A_{\a_\ell\rangle}
\nonumber\\
&&{} -\ell\omega^{\nu}\ep_{\nu\alpha( \a_\ell} F_{M_{\ell-1})}{}^{\alpha}
-{(\ell+1)(\ell+2)\over(2\ell+3)(2\ell+5)}E^{-(\ell+2)}
\left[E^{\ell+3}F_{\a\nu M_\ell}\right]'\sigma^{\a\nu }
\nonumber\\
&&{} -{2\ell\over (2\ell+3)}E^{-1/2}\left[E^{3/2}F_{\nu\langle
M_{\ell-1}}
\right]'\sigma_{\a_\ell\rangle}{}^{\nu} \nonumber\\
&&{}
 -E^{\ell-1}\left[E^{2-\ell} F_{\langle
M_{\ell-2}}\right]'\sigma_{\a_{\ell-1}\a_\ell\rangle}\,,
\label{r25}\end{eqnarray}
where a prime denotes $\p/\p E$. This is a fully nonlinear expression.

Multiplying Eq. (\ref{r25}) by $E^3$ and integrating over all
energies leads to the brightness multipole evolution equations:
\begin{eqnarray}
0 &=& \dot{\Pi}_{\langle M_\ell\rangle}+{{4\over3}}\Theta
\Pi_{M_\ell}+ \D_{\langle \a_\ell}\Pi_{M_{\ell-1}\rangle}
+{(\ell+1)\over(2\ell+3)}\D^{\nu} \Pi_{\nu M_\ell}
\nonumber\\
&&{} -{(\ell+1)(\ell-2)\over(2\ell+3)} A^{\nu} \Pi_{\nu M_\ell} +(\ell+3)
A_{\langle \a_\ell} \Pi_{M_{\ell-1}\rangle}\nonumber\\
&&{} -\ell\omega^{\nu}\ep_{\nu\alpha( \a_\ell}
\Pi_{M_{\ell-1})}{}^{\alpha}
 -{(\ell-1)(\ell+1)(\ell+2)\over(2\ell+3)(2\ell+5)}
\sigma^{\nu\alpha}\Pi_{\nu\alpha M_\ell} \nonumber\\
&&{} +{5\ell\over(2\ell+3)} \sigma^{\nu}{}_{\langle
\a_\ell} \Pi_{M_{\ell-1}\rangle \nu} -(\ell+2) \sigma_{\langle
\a_{\ell}\a_{\ell-1}} \Pi_{M_{\ell-2}\rangle}\,.
\label{r26}\end{eqnarray}
Once again, this is a fully nonlinear result.
The monopole evolution equation is just the energy conservation equation, i.e., Eq.~(\ref{e1i}) with $I=r$, and the dipole evolution equation is the momentum conservation equation~(\ref{e3i}), with $I=r$. The quadrupole evolution is given by Eq.~(\ref{nl8}).

\end{document}